\documentclass[%
reprint,
 amsmath,amssymb,
longbibliography,
 aps,
prb,
floatfix,
]{revtex4-2}

\usepackage{amssymb}
\usepackage[english]{babel}
\usepackage{graphicx}
\usepackage{multirow}
\usepackage[normalem]{ulem}
\usepackage{xcolor}
\usepackage{dcolumn}
\usepackage{color}
\usepackage{bm}
\usepackage{url} 
\usepackage{hyperref}
\hypersetup{
 colorlinks   = true,
 linkcolor=blue,
  citecolor    = blue,
  urlcolor=blue,
  } 
\usepackage{natbib}
\usepackage{filecontents}

\AtBeginDocument{%
    \newwrite\bibnotes
    \def\bibnotesext{mg_bibliography.bib}
    \immediate\openout\bibnotes=\jobname\bibnotesext
    \immediate\write\bibnotes{@CONTROL{REVTEX41Control}}
    \immediate\write\bibnotes{@CONTROL{%
    apsrev41Control,author="08",editor="1",pages="1",title="0",year="1"}}
     \if@filesw
     \immediate\write\@auxout{\string\citation{apsrev41Control}}%
    \fi
}%
      
\begin{document}

\title{Evidence of the disorder-independent electron-phonon scattering time in thin NbN films}

\author{A.\,I.\,Lomakin$^{1,2}$}
\author{E.\,M.\,Baeva$^{1,2}$}
\author{A.\,D.\,Triznova$^{2}$}
\author{N.\,A.\,Titova$^{2}$}
\author{P.\,I.\,Zolotov$^{2,3}$}
\author{A.\,V.\,Semenov$^{2}$}
\author{D.\,E.\,Sunegin$^{4}$}
\author{A.\,V.\,Lubenchenko$^{5}$}
\author{A.\,I.\,Kolbatova$^{1,2}$}
\author{ G.\,N.\,Goltsman$^{1,6}$}

\affiliation{$^1$National Research University Higher School of Economics, 20 Myasnitskaya St, Moscow, Russia\\
$^2$Moscow Pedagogical State University, 29 Malaya Pirogovskaya St, Moscow, Russia\\
$^3$ LLC Superconducting Nanotechnology (SCONTEL), 11A Derbenevskaya Naberezhnaya, Moscow, Russia\\
$^4$  Osipyan Institute of Solid State Physics RAS (ISSP RAS), 2 Academician Osipyan St, Chernogolovka, Moscow District, Russia\\
$^5$National Research University MPEI, 14 Krasnokazarmennaya St, Moscow, Russia\\
$^6$Russian Quantum Center, 100 Novaya Street, Skolkovo, Moscow, Russia}

\begin{abstract}
We report on experimental study of the effect of disorder on electronic parameters and inelastic scattering mechanisms in ultrathin superconducting NbN films, which are commonly used in single-photon detectors. An increase in disorder in the studied 2.5 nm thick NbN films characterized by Ioffe-Regel parameter from 6.3 to 1.6 is accompanied by a decrease in the critical temperature $T_c$ from 11.5 K to 3.4 K.
By measuring magnetoconductance in the range from $T_c$ to $\sim3T_c$, we extract the inelastic scattering rates of electrons, including electron-phonon (e-ph) scattering rates $\tau_{e-ph}^{-1}$. We observe that $\tau_{e-ph}^{-1}$ and their temperature dependencies are insensitive to disorder that is not described by the existing models of the e-ph scattering in disordered metals and can be due to the presence of weakly disordered metal grains. As the temperature decreases the temperature dependence $\tau_{e-ph}^{-1}$ changes from $T^3$ to $T^2$, which can be result of a decrease in the dimension of the phonons involved in the e-ph scattering. The obtained values of material parameters of ultrathin NbN films can be useful for optimization of performance of NbN-based electronic devices.

\end{abstract}

\maketitle
\section{INTRODUCTION}

Physical mechanisms governing the behavior of superconducting and electronic properties in ultrathin films have been studied extensively in order to understand the impact of disorder and quantum effects on electron transport in that sort of materials~\cite{Sacepe2020}. These fundamental studies are also motivated by usability of thin disordered films in superconducting devices, such as photon detectors~\cite{Semenov2021, Shurakov_2015, Zmuidzinas_2012, PdeVisser_2021}. To optimize the operation of these thin-film devices, it is essential to know parameters, which control the non-equilibrium response to radiation: for instance, electronic and phonon heat capacities, electron diffusivity, rates of inelastic electron-electron (e-e) and  electron-phonon (e-ph) scattering processes. Numerous studies of electron transport in disordered metals reveal significant impact of disorder on the mechanisms of inelastic scattering. For example, an enhancement of the e-e scattering rates is expected due to a strong elastic scattering of quasiparticles in thin disordered films~\cite{Altshuler1985} or due to the presence of a moderate density of magnetic impurities in low-dimensional devices~\cite{Anthore2003,Huard2005}. It is also proposed that strong disorder can modify the e-ph scattering, and one can expect weakening or strengthening of the e-ph interaction, depending on the specific properties of disordered systems~\cite{Schmid1973, Sergeev_Mitin2002}, or emergence of additional relaxation channels ~\cite{Shtyk2015}. In samples with reduced dimensions, relaxation processes also depend on samples size~\cite{Gershenson2001, Lin_2002, Karvonen2005}, which can lead to an even greater variety of effects in inelastic relaxation. Thus, an understanding of the role of disorder in inelastic scattering in thin-film devices can come mainly from an empirical study of a specific material.

Thin film of niobium nitride (NbN) is a typical material, in which disorder can be tuned from moderate to strong limit~\cite{Chockalingam2008, Chand2012, Zolotov2020}. This material has been extensively used for the production of modern electronic devices such as SNSPDs (Superconducting Nanowire Single Photon Detectors)~\cite{Goltsman2001,Divochiy2018}, HEB (Hot Electron Bolometer) mixers~\cite{Tretyakov2011}, microwave nanoinductors~\cite{Annunziata_2010} and resonators~\cite{Niepce2019}, quantum phase slip devices~\cite{Peltonen2013, Arutyunov_2016,Peltonen2016, Constantino2018, DeGraaf2018,Shaikhaidarov2022}, and etc. The choice of NbN for these technological applications has been justified by its relatively high superconducting critical temperature ($T_c\approx 10-12$\,K in $d \approx 5$\,nm thick films), high values of resistivity ($\rho > 100$\,$\mu\Omega$cm) and the relatively fast e-ph relaxation ($\tau_{e-ph} \sim 10$\,ps at $T_c\approx 10$\,K\cite{Gousev1994}). On the fundamental level there is observed an ongoing interest in effects of disorder on superconducting and normal state properties in NbN films~\cite{Chockalingam2008, Chand2012, Mondal2011, Delacour2012, Mondal2013, Noat2013, Carbillet2020}. Mechanisms of inelastic relaxation in thin NbN films have been studied with various experimental methods, however significant inconsistencies in the data have appeared in literature over the past decades~\cite{Gousev1994,Ilin2000, Zhang2003, Lin2013, Zhang2018, Sidorova2020,Zhang2020}. They are mainly caused by changing material parameters due to different deposition conditions. Nevertheless, a steady progress in production of NbN films~\cite{Linzen_2017,Knehr2021, Zolotov2020} opens up new horizons for further research, in particular for a systematic study of the effect of disorder on electron transport and inelastic scattering mechanisms in ultrathin films of NbN.

In this work we have prepared a set of ultrathin NbN samples with a tunable intrinsic disorder by depositing films in different conditions. Here we change only one deposition parameter at a time and fix the others.
We studied evolution of electronic parameters and inelastic scattering mechanisms with the increase of disorder. To study inelastic scattering we applied magnetoconductance measurements, which have been successfully used to study inelastic scattering in some samples of NbN~\cite{Shinozaki2019,Sidorova2020} and NbTiN ~\cite{Sidorova2021} recently. We have observed that the increase of disorder in the studied NbN films has no noticeable effect on the magnitude and the temperature dependence of the e-ph scattering rate $\tau_{e-ph}^{-1}$. The results for $\tau_{e-ph}^{-1}$ are also in good agreement with previous experimental data, extracted from the magnetoconductance for NbN films~\cite{Sidorova2020} as well as with the results obtained from a photoresponse of the NbN detector~\cite{Gousev1994}.

\section{SAMPLES AND MEASUREMENT SETUP}
Ultrathin NbN films are deposited using magnetron sputtering system (AJA International Inc.) with a background pressure of $9\times10^{-8}$\,torr. Samples are deposited on r-cut sapphire substrates by sputtering of Nb target in an argon-nitrogen atmosphere (99.998\% purity of both gases). Growth rate equal to $0.065$\,nm/s is controlled via quartz crystal microbalance in each deposition run. Studied films have thickness of $d = 2.5$\,nm. The level of disorder in five NbN films (s1-s5 in~\autoref{Table_1}) is varied by changing a substrate temperature in each deposition process $T_{dep}$: $500$, $400$, $300$, $150$, and $25$\,$^{\circ}$C (no additional heating), respectively. The films s1-s5 are also grown at a fixed nitrogen concentration of 22\%, while operating pressure is maintained constant at 3.6 mtorr. Heating of substrates was carried out by using the built-in resistive SiC heater with a PID controller. The most disordered samples are grown at the following conditions: $T_{dep}$ = 500\,$^{\circ}$C; 27\% of nitrogen at 6.5 mtorr for s6 and 23\% of nitrogen at 6.8 mtorr for s7. Operating pressure was adjusted using a throttle valve. As result, we obtained NbN films of the same thickness but of different resistance per square $R_s$ and critical temperature $T_c$ by varying heating of substrates and partial nitrogen pressure. To prevent unintentional oxidation of NbN films in the atmosphere, films are covered with a 5-nm thick passivating silicon layer \textit{in situ}. Structural characterization of a 2.5-nm thick NbN sample (s1) is performed using X-ray diffraction analysis, X-ray photoelectron spectroscopy (XPS) and atomic force microscopy (AFM). The details of structural characterization are given in Appendix A.

To study transport properties of NbN, we patterned films into 500-$\mu$m wide and 1000-$\mu$m long Hall-bars. Electrical transport measurements were carried out with Lake Shore 370 AC Resistance Bridge at a bias current less than 1\,$\mu$A. Normal-state resistance $R_s$ was measured in a four-probe configuration. The measurements were carried out in a homemade \textit{$^{4}$He} cryogenic insert immersed in a dewar and performed in a wide temperature range (from 300\,K to 1.7\,K). At low temperatures we measured magnetoresistance $R_s(B)$, temperature dependencies of the second critical magnetic field $B_{c2}(T)$ and the Hall constant $R_H^{25\,K}$ at 25\,K by applying perpendicular magnetic field $B$ up to 4\,T. By measuring $R(T)$-dependencies at different values of $B$ (not shown here) we determined the slope $dB_{c2}/dT$ at $T_{c}$. The latter allows to estimate the critical magnetic field $B_{c2}(0)$, the electron diffusivity $D$ and the Ginzburg-Landau (GL) coherence length $\xi_{GL}(0)$ using the following expressions $B_{c2}(0) = -0.69 T_c\left(dB_{c2}/dT\right)$, $D =-4k_B/(\pi e)\left(dB_{c2}/dT\right)^{-1}$, and $\xi_{GL}^2(0)=\pi \hbar D/\left(8k_B T_c\right)$. Here, the critical temperature $T_c$ is determined as temperature at $R_s = R_{max}/2$. 

\section{RESULTS AND DISCUSSION}

\subsection{Normal-state properties}
\begin{figure}[h!]
    \includegraphics[scale=1]{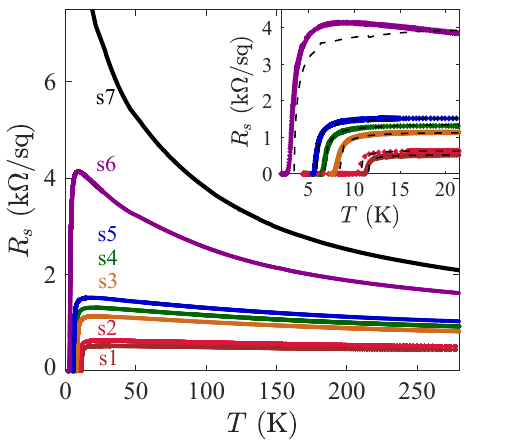}
    \caption{\label{fig_1} Temperature dependences of the sheet resistance $R_s(T)$ for all samples (main panel) and only for superconducting samples (s1-s6) in the narrow temperature range near their superconducting transitions (inset). The dashed lines represent fits obtained from the quantum corrections to conductivity taking into account superconducting fluctuations,  weak localization and e-e interaction (see text for details).}
\end{figure}

\autoref{fig_1} shows temperature dependencies of the sheet resistance $R_s$ for all NbN films under study. $R_s$ for all films gradually rises as temperature decreases and drops to zero at vicinity of $T_c$. The temperature trend of $R_s$ can be characterized by the resistance ratio parameter $r_R = R_{s}^{300 K}/R_{max}$, where $R_{max}$ is the maximal sheet resistance just above the resistive transition. Commonly observed $r_R<1$ is well-known behavior for disordered NbN films. It reflects sensitivity of electron system to quantum corrections, meanwhile contribution of e-ph scattering is considered to be negligible here. We also observed the suppression of $T_c$ with increasing disorder, which is  consistent with the fermionic mechanism of suppression of superconductivity in moderately disordered films (see Appendix B for details). 

\autoref{Table_1} gives an overview of metallic properties for studied NbN films. The details about the carrier density $n$, the Ioffe-Regel (disorder) parameter $k_Fl$, the mean free path $l$, the electron diffusivity $D$, and the density of states $N_0$ estimates are provided in Appendix C. High normal-state resistance $R_s$ and low resistance ratio parameter $r_R$ of the magnetron-sputtered NbN films are also known to be strongly influenced by grain-boundary scattering~\cite{Nigro1988,Tyan1994, Senapati2006}, which is related to electron transmission through grain boundaries~\cite{Reiss1986}. One should note that the graininess of ultrathin NbN films is not as pronounced as that of thick films (see AFM studies of ultrathin NbN films in Appendix A). Meanwhile, we assume that the electron transport through the grains can be considered as one of the main factors limiting the mean free path in NbN films (see Appendix C for details). 

\begin{table*}
\caption{\label{Table_1} Parameters of NbN films}
\begin{tabular}{c c c c c c c c c c c c c c c c}
\hline \hline
$\mathcal{N}\textsuperscript{\underline{o}}$
& $R_{s}^{300K}$& $r_R$ & $k_{F}l$&  $T_c$ &  $D$ & $\xi_{GL}(0)$ & $B_{c2}(0)$& $n$ $\times 10^{29}$& $\tau$ & $l$ &$N_0$ $\times 10^{28}$ & $R_{NS}$ & $\tau_{e-ph}(T_c)$ & $p$& $\varepsilon_F$\\

 & ($\Omega$/sq) & & & (K) & (cm$^2$/s)& (nm)& (T)& (m$^{-3}$) & (fs) & ($\mathrm{\AA}$)&  (eV$^{-1}$m$^{-3}$) & ($\Omega$/sq) & (ps) & & (eV) \\ \hline 
 s1 & 437 & 0.85 & 6.3 & 11.54 &0.59 &3.9& 14.8 & 1.9& 0.7&3.5 & 4.8& 497 & 8.3&3.05& 5.9 \\
s2 & 509 & 0.81 & 5.5& 10.76 &0.57 &4.0& 14.3& 1.8& 0.6& 3.16& 4.3 & 602 &7&3 & 6.2\\
s3 & 815 & 0.72 & 3.5& 8.43 & 0.36 &3.6& 17.9& 1.7& 0.4& 2& 4.3 & 1010 & 20.5&3.05 & 5.9 \\
s4 & 912 & 0.69 & 3.2 & 7.02 & 0.35 &3.8 & 15.3& 1.6& 0.35& 1.9 & 3.9 &1160 &21& 2.4 & 6.1\\
s5 & 1025 & 0.67 &2.8& 6.03 & 0.34 &4.1&13.5& 1.6 & 0.27& 1.65& 3.6 &1287 &20 &2 & 6.6\\
s6 & 1574 & 0.38 &2.1 &3.40  & 0.27 &4.9 &9.7&1& 0.28& 1.5 & 3.0 & 2550 &30 &1.7 & 5.0\\ 
s7 & 1950 & 0.26 & 1.6 & -  & 0.19 & - & -& 1.1& 0.15& 1.1 & 2.4 & - & 300 (T=1 K) & 1.7 &6.9  \\ \hline \hline
\end{tabular}
\end{table*}

\subsection{Magnetoconductance}\label{section3}

\begin{figure}[h!]
    \includegraphics[scale=1]{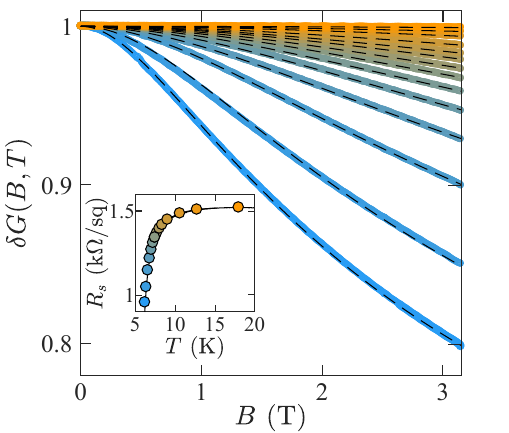}
    \caption{\label{fig_2}Experimental dependencies of the normalized magnetoconductance $\delta G(B,T)$ for a representative sample (s5). Different colors of the curves correspond to different operating temperatures marked on the $R_s(T)$-curve in the inset. The dashed black curves represent fits by Eq.~\eqref{fit_MC}.}
\end{figure}

To study inelastic relaxation in NbN films as a function of disorder, we measured magnetoresistance by varying magnetic field $B$ in the range from $0$ to $4$\,T at a set of fixed temperatures from $T_c$ to $30$\,K. The NbN films studied represent quasi two-dimensional (2D) system with respect to characteristic length scales, i.e. the thermal coherence length $L_T = \sqrt{\hbar D/k_B T}$ and the superconducting coherence length $\xi_{GL}$ ($d< \xi_{GL}, L_T$). The dimensionless change in magnetoconductance at the fixed $T$ can be determined from the measured $R_s(B,T)$ using the following expression:
\begin{equation}
\delta G(B,T) = \frac{2\pi^2 \hbar}{e^2}\left[R_s(B,T)^{-1}-R_s(0,T)^{-1}\right].
\label{Eq2}
\end{equation}
\autoref{fig_2} shows typical experimental dependencies of $\delta G(B,T)$ for NbN samples. The data, presented for a representative sample (s5), look similar for other samples.

Applying the fluctuation spectroscopy approach, we fit the experimental data in~\autoref{fig_2} by the relative magnetoconductance $\delta G^{QC}$ (see the Appendix D for details). In this approach $\delta G^{QC}$ for superconducting samples reflects the change in conductance originated from superconducting fluctuations and the weak localization. Using $T_{c}$ as a fitting parameter, we estimated the anomalous Maki-Thompson term, which contains information about the temperature-dependent electron phase-breaking rate $\tau_{\phi}^{-1}(T)$. Note that the choice of $T_c$ in 2D disordered superconducting films is ambiguous due to disorder induced spatial inhomogeneity of superconducting properties~\cite{Benfatto2009, Caprara2011,Baturina_2012}. Nevertheless, in our measurements the choice of $T_c$ barely affects the extracted values of $\tau_{\phi}(T)$ (see Appendix E for
details). The best fits of $\delta G(B,T)$ in~\autoref{fig_2} are shown with black curves with values of $T_{c}$ varying in under $2$\% in respect to the values determined at $R_s=R_{max}/2$. It is important to note that the phase-breaking length $L_{\phi}=\sqrt{D\tau_{\phi}}$ is larger than the film thickness $d$ in a considered temperature range. This fact supports validity of using 2D expressions for NbN films under study. The values of $\tau_{\phi}(T)$, obtained for the most disordered and non-superconducting sample (s7), are extracted by taking into account only the weak localization term (see Eq.~\ref{eq:fit_MC_1} in the Appendix D).

The same approach, with adding the e-e scattering term to the conductivity, can be exploited for fitting of $R_s(T)$-dependencies above $T_c$ (see the inset of \autoref{fig_1}). Here we use the expression $R_s(T)=1/(R_{NS}^{-1}+e^2/(2\pi^2\hbar)G^{QC}(0,T))$, where $R_{NS}$ is a fitting parameter and the values of $\tau_{\phi}$ in $G^{QC}(0,T)$ are those extracted from the magnetoconductance processing. The values of $R_{NS}$ turn out to be close to $R_s^{300 K}$ for moderately-disordered films. The discrepancy between $R_s^{300 K}$ and $R_{NS}$ in the strongly disordered NbN film (s6) may be explained by changing their granular properties (size of granules, intergranular properties) and approaching the Anderson-Mott transition~\cite{Mondal2011_anderson}.

\subsection{Electron phase-breaking rate}\label{section1}
\begin{figure}[h!]
    \includegraphics[scale=1]{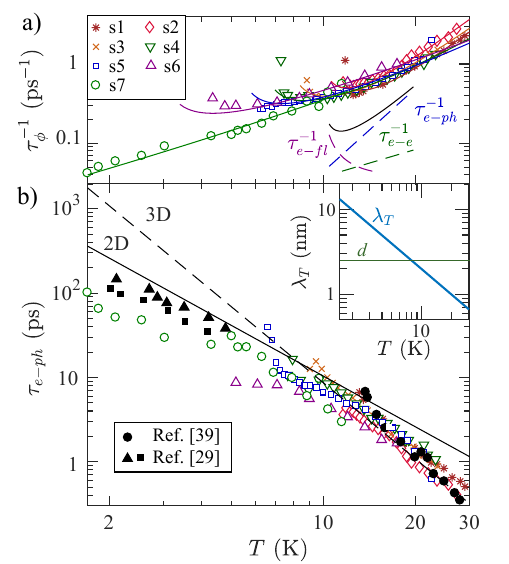}
    \caption{\label{fig_3}Temperature dependencies of (a) the electron dephasing rate $\tau_{\phi}^{-1}$ extracted from magnetoconductance measurements and (b) the e-ph scattering time $\tau_{e-ph}$ extracted from $\tau_{\phi}^{-1}$. The data is plotted in symbols on a log-log scale. In (a) the solid curves show the best fits of $\tau_{\phi}^{-1}$ by Eq.~\eqref{rates}. The dashed lines illustrate contributions of the e-ph, e-e, and SC dephasing mechanisms for sample s1. In (b) the solid and dashed lines demonstrates predictions for the e-ph scattering on three-dimensional (3D) and two-dimensional (2D) phonons, respectively, with the power-law dependences $\propto T^{-3}$ and $\propto T^{-2}$ in clean case. The inset in (b) presents the temperature dependence of the phonon wavelength and its comparison with thickness of NbN films studied in this work. } 
\end{figure}

As a next step, we explore the impact of disorder on inelastic scattering in NbN films.~\autoref{fig_3}(a) shows the temperature dependences of $\tau_{\phi}^{-1}$ for all studied samples. First of all, the data demonstrate the close resemblance of the results for NbN samples with different level of disorder. The data are characterized by close values of $\tau_{\phi}^{-1}$, as well as a similar power-law decrease in $\tau_{\phi}^{-1}$ with lowering temperature. The exact expression for $\tau_{\phi}^{-1}$ is represented by sum of scattering mechanisms due to superconducting fluctuations $\tau_{SC}^{-1}$, the e-e scattering rate $\tau_{e-e}^{-1}$, the spin-flip scattering rate $\tau_s^{-1}$, and the e-ph scattering rate $\tau_{e-ph}^{-1}$ as follows~\cite{Varlamov2018}:
\begin{subequations}\label{rates}
\begin{equation}
\tau_{\phi}^{-1}(T) = \tau_{SC}^{-1}+ \tau_{e-e}^{-1}+\tau_s^{-1}+\tau_{e-ph}^{-1}.
\tag{\ref{rates}}
\end{equation}
The presence of surface magnetic defects can significantly increase $\tau_{\phi}^{-1}$~\cite{Anthore2003, Huard2005} and lead to the $T$-independent behavior at low temperatures~\cite{Lin_2002,Huard2005}. The magnetic disorder in superconductors can lead to a time-reversal symmetry breaking caused by spin-flip scattering, altering superconducting state and suppressing $T_c$ more strongly than the nonmagnetic disorder~\cite{AbrikosovGorkov,Muller1971,Fominov2011}. Recent studies of Nb and NbN-based devices reveal unintentional surface magnetic disorder due to the unpaired spins in native oxide~\cite{Rogachev2005, Kumar2016,Samsonova2021, sheridan2021microscopic}. Nevertheless, we found that the passivating Si layer on top of NbN films fortunately prevented strong oxidation (see XPS analysis in Appendix A). In addition, the suppression of $T_c$ in our films can be explained by the strong electron interaction (see Appendix B). Thus, we treat the effects of magnetic disorder as negligible in analysis of $\tau_{\phi}^{-1}$-dependencies.

The e-e and superconducting fluctuations phase-breaking rates can be defined as~\cite{Altshuler1985,Brenig1985}:
\begin{align}
 \tau_{SC}^{-1} = \frac{\pi g k_B T}{\hbar} \frac{2\ln2}{\epsilon+\beta}, 
 \label{rate_fl} \\
\tau_{e-e}^{-1} =\frac{\pi g k_B T}{\hbar} \ln\left(\frac{1}{2\pi g}\right),
\label{rate_ee}
 \end{align}
 \end{subequations}
where $\beta=4\ln2/[\sqrt{\ln^2(1/(2\pi g))+64/(\pi^2 g)}+\ln(2\pi g)]$, $g = e^2R_s^{300K}/(2\pi^2 \hbar)$ and $\epsilon = \ln(T/T_{c})$. The value of $R_s^{300K}$ is taken here at the highest temperature of our measurements (300\,K), where the effect of e-e interactions is expected to be small~\cite{Khodas2003,Mondal2011_anderson}. The expression for the e-e scattering rate accounts only for the processes with small energy transfer - the so-called Nyquist quasielastic scattering, which dominates in our experimental temperature range with $T<\hbar/(k_B\tau)\sim10^3$\,K. The e-e scattering rate is expected to enhance greatly with increasing disorder~\cite{Altshuler1985}, but we estimate the increase of $\tau_{e-e}^{-1}$ as two times for the given change of $R_{s}^{300 K}$. Since $\tau_{\phi}^{-1}$ in our measurements is characterized by a stronger $T$-dependence than the e-e and SC phase-breaking rates, we believe that the e-ph scattering is the dominant dephasing process here. Applying that $\tau_{e-ph}=\alpha_{e-ph}^{-1}(T/T_{c})^p$, where $p$ and $\alpha_{e-ph}$ are fitting constants, we fit the data in \autoref{fig_3}(a) (see the dashed lines).~\autoref{Table_1} gives an overview of the best-fit values for the power index $p$ and $\alpha_{e-ph} = \tau_{e-ph}(T_c)$, where the latter reflects the e-ph scattering time at $T_c$.  

\autoref{fig_3}(b) shows the temperature dependencies of $\tau_{e-ph}$ extracted from the total dephasing rate $\tau_\phi^{-1}$ by subtracting $\tau_{e-e}^{-1}$ and $\tau_{e-sf}^{-1}$. We observe that the magnitude and the temperature dependences of $\tau_{e-ph}$ for the studied NbN films do not depend on disorder, but demonstrate the non-monotonic temperature dependence: it is proportional to $T^{-3}$ above $10$\,K, and it modifies to $T^{-2}$ at lower temperatures. To compare our findings with the previous results, we added to \autoref{fig_3}(b) the experimental data for some different NbN films measured by magnetoconductance in Ref.~\cite{Sidorova2020} and extracted from a response to the amplitude-modulated radiation (AMAR method) of NbN detector in Ref.~\cite{Gousev1994,EPH_vs_eph}. We observe the close agreement between all experimental data for $\tau_{e-ph}$ obtained for various NbN films. Below we will discuss potential mechanisms responsible for the observed rates of the e-ph scattering.

\subsection{Electron-phonon scattering in ultrathin films}
In search of potential explanation for the observed $\tau_\phi^{-1}$ behavior, we turned to the existing models of the e-ph scattering in thin disordered metals~\cite{Pippard1955,Schmid1973,Rammer1986}. In general, the e-ph coupling occurs because the passing phonons distort the local lattice structure and conduction electrons respond to the resulting band distortion. A widely used standard model of the low temperature e-ph scattering in bulk metals (i.e. jellium model)~\cite{Gantmakher_1974,Wellstood1994} assumes (i) a clean three-dimensional (3D) free electron gas with a spherical Fermi surface; (ii) a Debye description of the acoustic phonons; (iii) the scalar deformation potential, expected to be dominant at long-wavelength phonons and (iv) the dimensions of the metal are much longer than the average phonon wavelength (3D phonon spectrum). In this model, the average e-ph scattering rate appropriate for these assumptions is given by ~\cite{Gantmakher_1974,Wellstood1994}: $\tau_{e-ph}^{-1} = [6\zeta(3)(k_B T)^3(2\varepsilon_F/3)^2]/[2\pi\rho_m\Omega\hbar^4u_s^4v_F]$, where $\hbar$ is the reduced Planck's constant, $\rho_m$ the mass density, $\Omega$ the metal volume, $u_s$ the sound speed, $v_F$ the Fermi velocity, and $\varepsilon_F = 3n/(2N_0)$ the Fermi energy. In disordered metals the e-ph scattering is non-local with a characteristic size of the interaction region about the phonon wavelength $\lambda_T$. In the diffusive limit, when $l\ll \lambda_T$, the theory considers the following modifications of the rate $\tau_{e-ph}^{-1}\propto T^{4}l$~\cite{Pippard1955,Schmid1973,Rammer1986} and $\tau_{e-ph}^{-1}\propto T^{2}l^{-1}$~\cite{Sergeev_Mitin2002}, depending on the dominant phonon polarization (longitudinal or transverse ones) and type of disorder (vibrating or static types). One should note that numerous studies of the inelastic scattering with magnetoconductance reveal that there does not exist a universal temperature behavior of $\tau_{e-ph}^{-1}$ in disordered conductors~\cite{Lin_2002}. In particular, the value of the temperature exponent $p$ might be quite sensitive to the microscopic quality and the intrinsic material properties such as characteristic of the Fermi surface~\cite{Prunnila2005, Qu2005}, the dimensionality of the electron and phonon systems~\cite{Sergeev2005}, nontrivial phonon dispersion~\cite{Ono_2020}, etc.  

Study of the inelastic scattering with magnetoconductance in NbN films reveal that the power-law index $p$ and the magnitude of $\tau_{e-ph}^{-1}$ do not change with increasing disorder in explicit way. Note that the spread of the estimated values of the Fermi energy (see Table~\ref{Table_1}) is within 10\%, so in the further analysis, we assume that the change in the electronic parameters makes a negligible contribution to the change in the e-ph scattering. On the other hand, one would expect that $\tau_{e-ph}^{-1}$ strongly depends on the phonon properties, which, in samples with reduced dimensions, may differ from the Debye spectrum, accepted in the models. Previous studies of the response of thin NbN films to the modulated terahertz radiation revealed a $T^{1.6}$-dependence of the e-ph relaxation rate~\cite{Gousev1994} (see the black-filled symbols in \autoref{fig_3}(b)), which has been explained by the renormalization of the phonon spectrum in thin films.

In this work, we compare the experimental data in \autoref{fig_3}(b) with the theoretical predictions for the e-ph scattering in clean case ($\lambda_T < l$). One can describe the electron scattering on 3D phonons by the expression $\tau_{e-ph, 3d}^{-1} = (7 \pi\zeta(3)\lambda_{3d} k_B T^3)/(2\hbar \theta_{D,3d}^2)$~\cite{Pethick_1979}, and for 2D phonons by $\tau_{e-ph, 2d}^{-1} = \lambda_{2d} k_B T^2/(\hbar \theta_{D,2d})$. Here $\lambda$ is the e-ph coupling constant, $\theta_{D,3d} = \hbar(6\pi^2)^{1/3}u_{3d}/(a k_B)$ and $u_{3d} = [1/3(2/u_t^3+1/u_l^3)]^{-1/3}$, $\theta_{D,2d} = \hbar(4\pi)^{1/2}u_{2d}/(a k_B)$ and $u_{2d} = [1/2(1/u_t^2+1/u_l^2)]^{-1/2}$ are the Debye temperatures and the mean sound velocities, respectively, for 3D and 2D phonons, $a=0.44$\,nm is the NbN lattice constant. The fitting values $\lambda_{3D}=0.98$ (the same value as in Ref.~\cite{Babu_2019}), $\theta_{D,3d}=116$\,K with $u_{3d} =1.73\times 10^3$\,m/s for 3D case and $\lambda_{2D}=1$, $\theta_{D,2d}=118$\,K with $u_{2d} =1.93\times 10^3$\,m/s for 2D case provide the close agreement with the experimental data in \autoref{fig_3}(b). Note that the fitting value of $\theta_{D,3d}$ is slightly lower the estimated values previously reported for thin NbN films ($\theta_D\sim 172-174$\,K)~\cite{Chockalingam2008,Sidorova2020}. Taking into account the average value of $u_m$, one can estimate the phonon wavelength $\lambda_T= h u_m/(2.82 k_BT)$ (here $2.82$ is a constant given in the dominant phonon approximation), which is comparable with the film thickness $d$, see the inset of \autoref{fig_3}(b). Here the crossover $\lambda_T\approx d$ is expected at $T_{cr}\approx 10$\,K. Therefore, the e-ph scattering in ultrathin NbN films can be considered in 3D regime at temperatures higher than $T_{cr}$, and in 2D regime at low temperatures.  

To ensure the condition that the film is two-dimensional for the phonons, one can assume that there is a substantial acoustic mismatch between the film and and the substrate~\cite{Kaplan1979, Sidorova2018}. In that case, the quantization of the phonon spectrum in direction perpendicular to the film may be significant, leading to reduced phonon density of  states and, hence, a weakened temperature dependence. Using the fitting parameters $\theta_{D} = 116$\,K, $u_{m} = 1.73\times 10^3$\,m/s, and the value of the phonon escape time $\tau_{esc}= 120$\,ps (see Appendix F for details), we estimate the phonon transmission coefficient $\eta = 4d/(\tau_{esc}u_{m})\approx 0.05$, which is consistent with the previous results for NbN/Al$_2$O$_3$ interface~\cite{Sidorova2020}. This small value of $\eta$ indicates that available directions for the phonon escape from the ultrathin film are strongly restricted.

Another surprising fact is that the observed T-dependences of $\tau_{e-ph}$ for NbN films are close to predictions of the simple model for the e-ph scattering in clean metals. Typically, NbN films have a polycrystalline structure~\cite{Lin_2013}, which can be considered as a composite of crystalline grains separated by thin amorphous boundaries ~\cite{Zhang7735}. In our study, we observe a significant increase of the sheet resistance with change of deposition conditions, which may indicate a change in crystalline properties. At the same time, we see that the disorder weakly affects the e-ph scattering rates in these NbN films. One should note that the disorder in the models of the e-ph scattering in disordered metals~\cite{Schmid1973, Sergeev_Mitin2002,Lin_2002} is usually treated as point defects, while in NbN films the increase in resistance and decrease in the mean free path may be related to a different origin, for instance, electron tunneling through barriers, which separate NbN grains. Thus, the disorder-independent character of $\tau_{e-ph}^{-1}$ in granular samples might come from the presence of weakly disordered metal grains. However, this assumption requires further theoretical and empirical study.

\section{Discussion}
In this paper, we investigate the evolution of electronic parameters and inelastic scattering rates with increasing disorder in ultrathin NbN films. We find that the inelastic scattering rates of electrons and their temperature dependencies are close for NbN films of different microscopic quality and with different levels of disorder. Our experimental results in \autoref{fig_3} are in agreement with previous studies of the inelastic scattering times in NbN films~\cite{Gousev1994, Sidorova2020}.

One should note that, in detectors, the energy relaxation time due to e-ph scattering $\tau_{E-PH}$ differ from $\tau_{e-ph}$ by a numerical coefficient $\tau_{E-PH} =\alpha \tau_{e-ph}$, where $\alpha\approx 0.6-0.1$ for $p = 2-4$, respectively~\cite{Ilin1998}. Thus, the weak dependence on disorder for $\tau_{e-ph}(T_c)$ in NbN films with a higher level of disorder and lower $T_c$ values can facilitate for increase of sensitivity and spectral characteristics of HEBs devices~\cite{Shurakov_2015} as well as for increase the efficiency of single photon detection of SNSPDs. In the latter case, the fast inelastic scattering and the electron diffusivity both affect the hotspot size and the superconducting gap suppression~\cite{Vodolazov2017}. This expectation correlates with a recent breakthrough in detection efficiency in SNSPDs, when one uses NbN films with high normal-state resistance values~\cite{Hofher2010, Zolotov2020, Korneeva_2021}. On the other hand, one should keep in mind that the operation of the detectors is also limited by the phonon escape rate, which is proportional to the sound velocity. As shown in Section~D, the analysis of $\tau_{e-ph}^{-1}(T)$-dependence indicates low values of the sound velocity, which in turn can affect the phonon escape rate and limit the device response. Thus, further experimental study of the features of phonon transport in ultrathin disordered metals is advisable.     

In summary, we observe the disorder-independent character of the electron-phonon scattering time in ultrathin NbN films. The observed results are not described by existing models of the e-ph scattering in disordered metals. The experimental values of $\tau_{e-ph}^{-1}$, measured above $10$\,K, are proportional to $T^3$, expected for the electron scattering on three-dimensional acoustic phonons in clean case. At lower temperatures, $\tau_{e-ph}^{-1}$ modifies to $T^2$, which is likely due to lowering the dimensionality of the phonons involved in the e-ph scattering. Our results call for further theoretical and experimental studies of the e-ph scattering in the presence of strong disorder. 

\section*{Acknowledgement}
We acknowledge valuable discussions with V.S. Khrapai and M. Sidorova. We are grateful to S.V. Simonov for their assistance with the X-ray studies. X-ray studies were performed using Rigaku SmartLab SE diffractometer at shared facility center of ISSP RAS. The growth of NbN films, material characterization and magnetotransport measurements were funded by the RFBR project No. 19-32-60076. The surface analysis and analysis of transport properties were carried out with the financial support of the Ministry of science and higher education of the Russian Federation in the framework of the Agreement No. 075-11-2022-026 dated 06.04.2022. 

\section*{\label{ap1} Appendix A: X-ray diffraction, XPS analysis and atomic force microscopy of a NbN film}

\begin{figure}[h!]
    \includegraphics[scale=1]{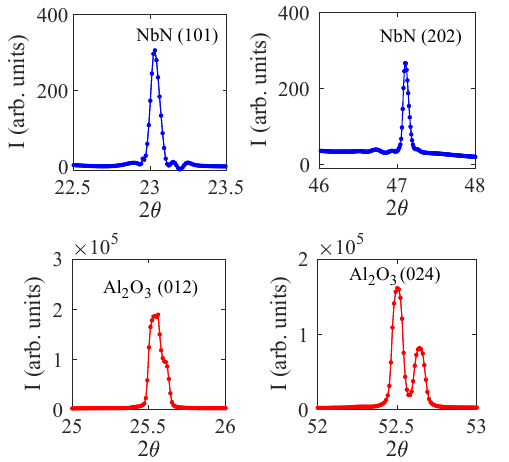}
    \caption{\label{fig_4} X-ray diffraction study of NbN sample (s1). Coupled $\omega - 2\theta$ scan, obtained with a RIGAKU diffractometer using CuK$\alpha1$ source. The positions of diffraction peaks with the corresponding diffraction angles are the followings: (a) 23.03$^{\circ}$ - NbN (101); (b) 47.01$^{\circ}$ - NbN(202); (c) 25.55$^{\circ}$ - Al$_2$O$_3$ (012) and (d) 52.57$^{\circ}$ - Al$_2$O$_3$ (024). The XRD data indicates that the studied sample has a tetragonal phase ($\gamma$-phase) with the space group I-4m2.}
\end{figure}

Structural characterization of a 2.5-nm thick NbN sample (s1) is shown in~\autoref{fig_4}. We plot the X-ray diffraction (XRD) data for a coupled $\omega - 2\theta$ scan and identify main diffraction peaks of NbN (101), (202) and Al$_2$O$_3$ (012), (024) for film and for substrate, respectively. The x-ray data indicates that NbN$_x$ has a tetragonal phase ($\gamma$-phase), which is generally characterized by $x = 0.75 -0.84$~\cite{Farha2013}. This phase is considered to be a deformed cubic phase ($\delta$-phase) with a similar transition temperature and the electronic spectrum as in a $\delta$-phase but has less N content~\cite{Ethridge1996, Pan_2021}. The observed $\gamma$-phase is in accordance with the phase diagram of NbN calculated as a function of the substrate heating temperature~\cite{Scheerer1979}. 

\begin{figure}[h!]
   \includegraphics[scale=1]{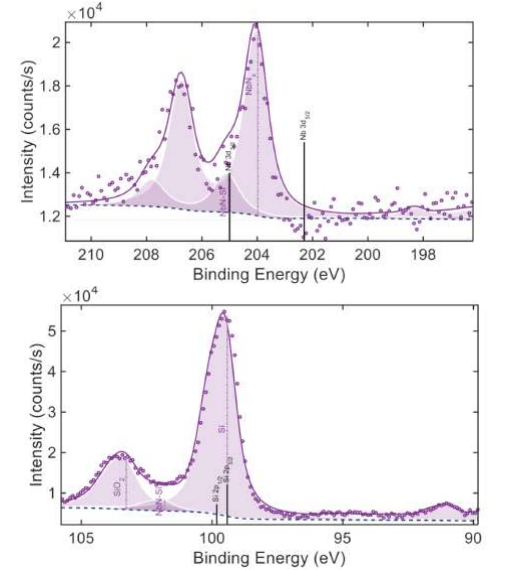}
    \caption{\label{fig_5} XPS spectra of  NbN sample (s1). The detailed scans of strong lines: lines Nb 3d (top figure) and lines Si 2p (bottom figure). The circles show the recorded detailed spectra, the solid lines are calculated using method described in Ref.~\cite{LUBENCHENKO2018711}, the area show separate calculated peaks. }
\end{figure}

\autoref{fig_5} displays the experimental data on X-ray photoelectron spectroscopy (XPS) presented for the least-disordered  sample in the set (s1). The XPS studies of samples surfaces were performed with the help of the electron-ion spectroscopy module based on Nanofab 25 (NT-MDT) platform in the analysis chamber an ultrahigh oil-free vacuum about $10^{–6}$\,Pa. \autoref{fig_5} shows decomposition of XPS lines of an element of interest (Nb, N, O, Al) into component peaks that reveals presence of different phases in the sample (NbN$_x$, NbN-Si).The XPS result are obtained on the base of the method described in Ref.~\cite{LUBENCHENKO2018711}.  The results allow to extract thicknesses of layers of different phases. The data for the studied NbN sample is presented in \autoref{Table_2}.

\begin{table}
\caption{\label{Table_2} Chemical and phase depth profile of the ultra-thin NbN film (s1).}
\begin{tabular}{c c c }
\hline \hline
& Formula & $d$ (nm) \\ \hline 
1 & SiO$_2$ & 1.03$\pm$0.19 \\
2 & Si & 5.3$\pm$0.5 \\
3 & NbN-Si & 0.59$\pm$0.11 \\
4 & NbN & not records \\
Substrate & Al$_2$O$_3$ & Inf \\\hline \hline
\end{tabular}
\end{table}

\begin{figure}[h!]
    \includegraphics[scale=1]{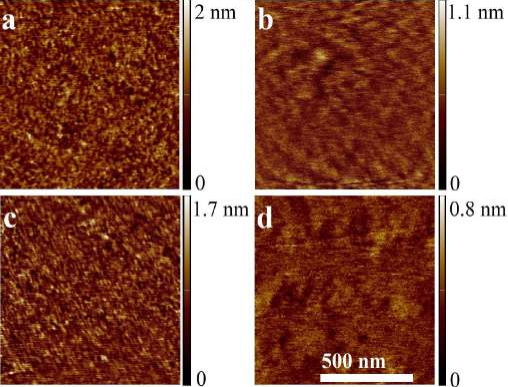}
    \caption{\label{fig_6} AFM images of representative NbN samples and a substrate: (a) s1, (b) s5, (c) s6, (d) r-cut sapphire. The rms roughness of the samples (s1, s5, s6) and the substrate are 0.20\,nm, 0.13\,nm, 0.19\,nm, and 0.06\,nm, respectively.}
\end{figure}

Figure~\ref{fig_6} shows the atomic force microscope (AFM) images of a few NbN samples covered by 5-nm thick Si layer and a sapphire substrate. The AFM images are obtained with NT-MDT (INTEGRA series) setup in classical contact mode. The rms surface roughness of NbN films is below 0.2\,nm, which corresponds to atomically smooth surface. In contrast to thick NbN films ($d>10$\,nm), which have a pronounced granular structure~\cite{Soldatenkova2021}, ultrathin NbN can be considered as a quasiamorphous matrix with the grain size smaller the film thickness. 

Surface topography of NbN$_x$ is known to be strongly affected by the substrate heating temperature $T_{dep}$ and the nitrogen partial pressure during deposition process~\cite{Benkahoul2005}. As shown in Figure~\ref{fig_6}, the microstructure of the samples deposited at high $T_{dep}$ (s1, s6) is compact-grained, since heating of the substrate increases mobility of adatoms and increases density of the film. The sample deposited at room temperature (s5) is characterized by less compact microstructure without pronounced grains. As the nitrogen content in the system changed, various and interrelated types of crystallographic structures are expected to form~\cite{Scheerer1979,Benkahoul2005}. With increase of the nitrogen background pressure, one can expect that the kinetic energy and flux of the ablated materials are reduced, and the latter results in less nitrogen ratio in NbN$_x$ film as well as a decrease in material density~\cite{Farha2013}. 

\section*{\label{ap2} Appendix B: variation of $T_c$ in ultrathin NbN films}

\begin{figure}[h!]
    \includegraphics[scale=1]{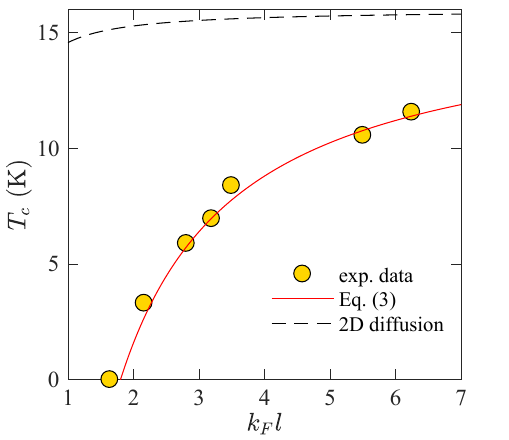}
    \caption{\label{fig_7} Suppression of the critical temperature $T_c$ with disorder. The experimental data (symbols) are in comparison with the fermionic mechanism of $T_c$ suppression (see the red solid line). The contribution of the 2D diffusion motion of electrons in Eq.\eqref{Eq1} is shown by the black dashed line.}
\end{figure}

Here we discuss the correlation between $T_c$ and $k_Fl$, shown in~\autoref{fig_7}. The breakdown of the superconductivity with the increase of disorder can be related to two different, but not mutually exclusive effects~\cite{Sacepe2020}. The first effect results from a decrease of electronic screening, which enhances the e-e repulsion and partially cancels the e-ph mediated attractive interaction. This fermionic effect leads to a mechanism described in Ref.~\cite{Finkelstein94,Antonenko2020}. The second effect comes from the disorder-induced decrease in superfluid density, which makes a superconductor susceptible to phase fluctuations~\cite{Raychaudhuri_2021, Dutta_2022}. The latter is known as a bosonic mechanism.

As shown in~\autoref{fig_7}, the suppression of $T_c$ follows the expectations of the fermionic mechanism, which can be expressed in the following form~\cite{Antonenko2020}:
\begin{equation}
\frac{T_{c0}-T_c}{T_{c0}} = \frac{\alpha_{3D}}{k_Fl}+\frac{\lambda_{ee}}{2(k_Fl)(k_F d)} \ln^3 \frac{\hbar}{T_{c0}\tau_D},
\label{Eq1}
\end{equation}
where the first term in rhs corresponds to the three-dimensional ballistic motion of electrons, the second term in rhs is related to 2D diffusive motion of electrons. Here $T_{c0}$ is the critical temperature in bulk, $\lambda_{ee}$ is the electron–electron coupling constant, and $\alpha_{3D}$ is the material dependent coefficient of the order of unity. The experimental data is in good agreement with Eq.~\ref{Eq1} taking $\lambda_{ee}=1/2$, $k_F=1.6\times 10^{10}$\,m$^{-1}$, $\tau_D$=40\,fs and treating $T_{c0}= 16$\,K and $\alpha_{3D}=1.8$ as the fitting parameters, see the red solid line in~\autoref{fig_7}.
It is instructive to note that suppression of $T_c$, related to 2D diffusive nature of electron motion, is weak here (the black dashed line), and it is likely determined by the 3D ballistics~\cite{Antonenko2020}. The observed dependence can be associated with the following hierarchy of length scales $l\ll d\simeq \xi$ in the studied NbN films, and thus the suppression of $T_c$ is likely to be controlled by $k_Fl$ parameter rather than $R_s$. One should also note that the change in the density of states at the Fermi level is small (see values of $N_0$ in \autoref{Table_1}) and cannot distort the analysis within the framework of the fermionic scenario.

Meanwhile, in the previous studies \cite{Mondal2011,Chand2012}, the electronic system in NbN films was observed to undergo the fermionic scenario at moderate level of disorder ($k_Fl>3$) and cross to a bosonic route at stronger disorder ($k_Fl<3$). The estimated level of disorder in our NbN films, in comparison to those in Ref.~\cite{Chand2012}, shows that the e-e repulsion can play a major role in destruction of the superconducting state for samples s1-s5, whereas the strong phase fluctuations due to small superfluid density can be more pronounced for the most disordered samples in this study (s6 and s7). The latter means that these two effects, which provide the disorder-induced suppression of $T_c$, should be considered collectively. However, we leave further refinements beyond the scope of this work, since the exact microscopic picture of the suppression of $T_c$ does not play a large role in the results obtained for $\tau_{\phi} (T)$ (see Appendix E for details).

\section*{\label{ap3} Appendix C: electron transport}
\par We calculate all electronic parameters at the highest temperature of our measurements (300\,K), where the effect of e-e interactions is expected to be small~\cite{Khodas2003}. Firstly, to estimate the Hall coefficient $R_{H}^{300 K}$ we consider the expression $\left(R_{H}^{25 K}-R_{H}^{300K}\right)/R_{H}^{300K}=\gamma \left(R_{s}^{25K}-R_{s}^{300K}\right)/R_{s}^{300K}$~\cite{Altshuler1985}, where $\gamma=0.68$ is an empirical parameter~\cite{Madhavi2009}. Using the value of $R_{H}^{300 K}$ we derive the carrier density $n = - B/(edR_{H}^{300 K})$, where $e$ is the electron charge. It is also important to note that samples s1-s5 are characterized by a moderate change ($<20$\,\%) in the carrier density $n$ at increase of $R_{s}^{300 K}$ in two times. Meanwhile, the most disordered samples (s6 and s7) are characterized by a twofold decrease of $n$ at three- and fourfold increase of $R_{s}^{300 K}$. The latter change in $n$ can be a result of the increase of Nb and N vacancies due to the increase of the nitrogen partial pressure in the mixture during deposition of NbN~\cite{Chockalingam2008}.

\begin{figure}[h!]
    \includegraphics[scale=1]{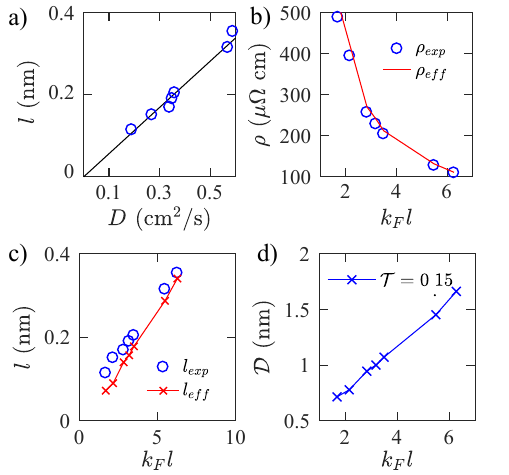}
    \caption{\label{fig_8}
    Electron transport in the normal state. (a) The relationship of the electron diffusivity $D$ and the mean free path $l$. The solid line shows the approximation with $l=3D/v_F$, where $v_F\approx (5.3\pm 0.3)\times 10^5$\,m/s. (b-d) Grain boundary scattering. In (b) $\rho_{300K}$ and $\rho_{eff}$ vs. the disorder parameter $k_Fl$. (c) Experimental values of mean-free path $l_{exp}$ and $l_{eff}$ as function of disorder. (d) The dependence of the effective grain size $\mathcal{D}$ on disorder $k_Fl$ at the fixed value $\mathcal{T}\approx 0.15$~\cite{Senapati2006}. }
\end{figure}

To characterize electron transport, we estimate the mean free path $l$, the elastic scattering time $\tau$ and the single-spin density of states (DOS) at the Fermi level $N_0$ assuming the free electron model~\cite{Kittel2004} and using the following expressions $l= \hbar k_F/(e^2n R_{s}^{300K}d)$, $\tau = l^2/3D$, and $N_0 = 1/(2 R_{s}^{300K}d e^2 D)$, where $k_F = (3\pi^2n)^{1/3}$ is the Fermi wavevector. We believe that this \textit{ab-initio} estimates are well justified, since $l\approx 1-3.5$\,${\AA}$, which is in order of the lattice constant in our disordered NbN films, thereby a possible Fermi surface anisotropy can be negligible. The electronic transport is also characterized by the Ioffe-Regel parameter $k_Fl$, which is a common indicator of the disorder in homogeneously disordered material. All estimated parameters are listed in~\autoref{Table_1}. In addition, \autoref{fig_8}(a) shows the relationship between the experimental values of $l$ and $D$, which can be approximated as $l=3D/v_F$, where $v_F\approx (5.3\pm 0.3)\times 10^5$\,m/s is the Fermi velocity determined as a fitting parameter here. The estimate of $v_F$ is two times lower than the one previously reported~\cite{Mondal2013}, and it corresponds to the effective mass of charge carriers $m_{eff}=\hbar k_F/v_F$, which decreases from $4.1m_e$ to $3.3m_e$ as $R_{s}^{300 K}$ increases (here $m_e$ is the mass of a free electron). Note that taking into account these values of $m_{eff}$ allows to correct underestimation of $n$ determined from optical measurements in NbN~\cite{Semenov_2009}.

In an attempt to understand electron transport in NbN films better, we performed a simple analysis of resistivity in terms of the grain boundary model~\cite{Reiss1986,Nigro1988,Tyan1994,Senapati2006}. In general, resistivity of granular films can be qualitatively given in the following way~\cite{Reiss1986}: $\rho_{eff} = (3\pi^2)^{1/3}\hbar/(n_0^{2/3} e^2l_0)\mathcal{T}^{-l_0/\mathcal{D}}$, where $l_0$ is the mean free path in absence of granularity, $\mathcal{D}$ is the mean grain size, $\mathcal{T}$ is the transmission probability of electrons through the grain boundary, $n_0$ is the carrier density of bulk material without crystal boundaries. Resistance here is increased since the effective mean free path is reduced due to the reflection on the grain boundaries: $l_{eff} = l_0\mathcal{T}^{l_0/\mathcal{D}}$. ~\autoref{fig_8}(b-d) demonstrates an attempt to evaluate the effect of granularity on resistivity of the studied NbN films. In ~\autoref{fig_8}(b-c) we compare the values of resistivity and the mean free path deduced from transport measurements with $\rho_{eff}$ and $l_{eff}$ as a function of disorder ($k_Fl$). Taking into account $n_0\approx2.29\times10^{23}$\,cm$^{-3}$ and $l_0\approx 1.1$\,nm reported for crystalline NbN films~\cite{Yan2018}, we evaluated $l_{eff}$, which is a function of the coupled parameters $\mathcal{T}$ and $\mathcal{D}$. In contrast to thick NbN films, which exhibit a pronounced polycrystalline structure~\cite{Soldatenkova_2021}, ultrathin NbN films can represent a quasiamorphous matrix with the expected grain size smaller than the film thickness but larger than the characteristic unit cell. Taking into account the fixed value $\mathcal{T}\approx 0.15$~\cite{Senapati2006}, we estimate the effective size of a crystalline grain $\mathcal{D}$ in studied NbN samples (~\autoref{fig_8}(d)), which falls within the range $a<\mathcal{D}< d$, where $d=2.5$\,nm is the film thickness and $a\approx0.44$\,nm is the lattice constant~\cite{Oya1974,Babu_2019}. As revealed in AFM studies (see Appendix A), ultrathin NbN films grown at different substrate temperatures are characterized by different grain packing density: NbN films deposited at high $T_{dep}$ typically have a more compact and dense microstructure than films deposited at low $T_{dep}$. In case of films deposited at different nitrogen partial pressures, this difference is not so obvious. Meanwhile, one can expect a decrease in the effective grain size for NbN films deposited at lower $T_{dep}$, as well as under excessive nitrogen partial pressure ~\cite{Farha2013}. 
Thus, increase of disorder in NbN films may reflect change of microstructure due to emergence of some granular structure as well as an increase in the number of point scatterers, such as vacancies at lattice sites.

\section*{ \label{ap4} Appendix D: Magnetoconductance}
To derive $\tau_{\phi}$ from the magnetoconductance at temperatures $T\geq T_c$, we fit data with the dimensionless change in magnetoconductance, which is given by expression:
\begin{subequations}\label{fit_MC} \begin{equation}
\tag{\ref{fit_MC}}
\delta G^{QC} = G^{QC}(B,T)-G^{QC}(0,T),
\end{equation}
where $G^{QC}(B,T)$ and $G^{QC}(0,T)$ are a sum of four terms of quantum corrections to conductivity at finite and zero magnetic fields: the weak localization (WL), the Aslamazov-Larkin (AL) term, the density of states (DOS) contribution term, and Maki-Thomson (MT) term.\\ The sum of these terms is given by the following expression~\cite{Varlamov2018, Glatz2011,Lopes1985, Rosenbaum1985}:
\begin{align}
G^{QC}(B,T) =\underbrace{ \frac{\pi^2\epsilon}{4h^2}\left[\psi\left(\frac{1}{2}+\frac{\epsilon}{2h}\right)-\psi\left(1+\frac{\epsilon}{2h}\right)+\frac{h}{\epsilon}\right]}_{\rm{AL}}\nonumber\\
\underbrace{-\frac{28\zeta\left(3\right)}{\pi^2} \left[\ln \left(\frac{1}{2h}\right) -\psi\left(\frac{1}{2}+\frac{\epsilon}{2h}\right)\right]}_{\rm{DOS}}\nonumber\\
-\underbrace{\beta_{MT}(T,\tau_\phi) \left[\psi\left(\frac{1}{2}+\frac{B_\phi}{B}\right)-\psi\left(\frac{1}{2}+\frac{B_\phi}{B}\frac{\epsilon}{\gamma_{\phi}}\right)\right]}_{\rm{MT}}\nonumber\\
\underbrace{+\frac{3}{2}\psi\left(\frac{1}{2}+\frac{B_2}{B}\right)-\psi\left(\frac{1}{2}+\frac{B_1}{B}\right)-\frac{1}{2}\psi\left(\frac{1}{2}+\frac{B_3}{B}\right)}_{\rm{WL}}
	\label{eq:fit_MC_1}
\end{align}
where $\psi(x)$ is Digamma function, $h = 0.69 B/B_{c2}(0)$ and $\epsilon = \ln(T/T_{c})$, $\gamma_{\phi} = \pi\hbar/(8k_BT\tau_{\phi})$, $B_1 = B_{0}+B_{so}$, $B_2 = B_{\phi}+4B_{so}/3+2B_{s}/3$, and $B_3 = B_{\phi}+2B_{s}$. The characteristic fields are defined as $B_{0} = \hbar/(4eD\tau)$, $B_{so} = \hbar/(4eD\tau_{so})$, and $B_{\phi} = \hbar/(4eD\tau_{\phi})$, where $\tau$, $\tau_{so}$, and $\tau_{\phi}$ are the relaxation time for elastic, spin-orbit, and phase-breaking scattering, respectively. The spin-orbit time can be roughly estimated using $\tau_{so}=\tau(\alpha Z)^{-4}\approx0.01 - 0.1$\,ps, where $\tau$ is the elastic scattering time, $\alpha$ is the fine structure constant, and $Z$ is the effective atomic number of material ($Z_\mathrm{NbN}\approx 24$). The magnetic scattering is assumed to negligible ($B_{s}=0$) since the samples are covered with a protective layer.
The coefficient in MT term is given as~\cite{Lopes1985}:
\begin{equation}
    \beta_{MT}(T,\tau_\phi) =\frac{\pi^2}{4}\sum_i (-1)^i \Gamma(\lvert i \rvert)-\sum_{i\geq 0}  \Gamma''(2 i+1) ,
\end{equation}
where $i=0,\pm 1,\pm 2...$ and $
    \Gamma^{-1}(\lvert i \rvert) =\epsilon +\psi\left(1/2+\lvert i \rvert/2 \right)-\psi\left(1/2\right) -\psi'\left(1/2+\lvert i \rvert/2\right)2\gamma_\phi/\pi^2$.
Note, that AL and DOS terms in the Eq.\eqref{eq:fit_MC_1} are asymptotics for low temperatures $\epsilon\ll 1$. Nevertheless, the derivation of $\tau_{\phi}$ at high temperatures do not dependent on presence of AL and DOS terms in the Eq.\eqref{eq:fit_MC_1}.\\
In the limit of zero magnetic field, Eq.\eqref{eq:fit_MC_1} with adding the e-e contribution~\cite{Altshuler1985} can be transformed to:
\begin{eqnarray}
G^{QC}(0,T) = \underbrace{\frac{\pi^2}{8\epsilon}}_{\rm{AL}}+\underbrace{\frac{28\zeta\left(3\right)}{\pi^2}\ln(\epsilon)}_{\rm{DOS}}+\underbrace{\beta_{MT}(T,\tau_\phi) \ln \left(\epsilon/\gamma_{\phi}\right)}_{\rm{MT}}\nonumber\\
\underbrace{+\ln\left(\frac{B_2}{B_1}\right)+\frac{1}{2}\ln\left(\frac{B_2}{B_3}\right)}_{\rm{WL}}+\underbrace{\ln\left(\frac{kT\tau}{\hbar}\right)}_{\rm{e-e}}
\label{eq:fit_MC_2}
\end{eqnarray}
\end{subequations}
The Eq.\eqref{eq:fit_MC_2} can be used for fitting $R_s(T)$-curve at low temperatures $\epsilon\ll 1$. To fit high-$T$ part of resistance we replace the AL and DOS terms with $2.32/\epsilon^3$ and $-\pi^4/96\epsilon^2$ asymptotics, respectively~\cite{Glatz2011}. 

\section*{\label{ap5}Appendix E: Influence of the choice of $T_c$ on estimates of $\tau_{\phi}$}

\begin{figure}[h!]
    \includegraphics[scale=1]{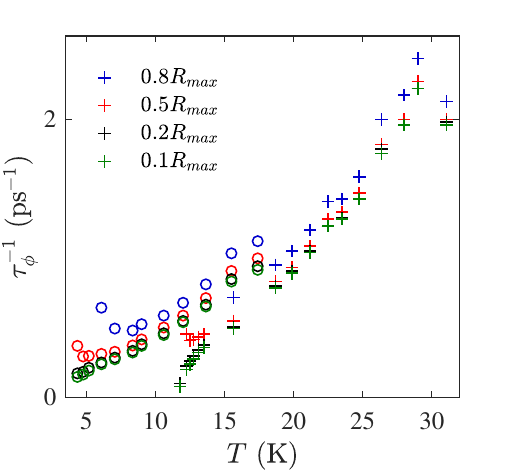}
    \caption{\label{fig_9} Temperature dependencies of the phase-breaking time $\tau_{\phi}(T)$ estimated at different points of the superconducting resistive transition, which correspond to different setpoints of $T_c$ ($0.8R_{max}$, $0.5R_{max}$, $0.2R_{max}$, and $0.1R_{max}$)}. The data are presented for two samples: s1 (cross-shaped symbols) and s6 (round symbols).
\end{figure}

To fit the experimental data for the magnetoconductance with the fluctuation spectroscopy, we use $T_c$ as a fitting parameter. Nominally, the critical temperature $T_c$ in our study is determined as the temperature at which $R_s = R_{max}/2$. \autoref{fig_9} shows that the choice of $T_c$ at different setpoints barely affects the extracted phase-breaking time $\tau_{\phi}$.  

\section*{\label{ap6}Appendix F: Estimate of the phonon escape time}

The characteristic phonon escape time can be found as $\tau_{esc} = (C_{ph}d)/(4\Sigma_{2D}T^3)$~\cite{Dane2022}, where $C_{ph}= 12\pi^4/5k_Ba^{-3}(T/\theta_D)^3$ is the phonon heat capacity in the 3D Debye model and $\Sigma_{2D}$ is the heat flow rate limited by the Kapitza resistance.
Taking into account the experimentally determined cooling rate for NbN-Al$_2$O$_3$ interface $\Sigma_{2D} \sim 120$\,WK$^{-4}$m$^{-2}$~\cite{Dane2022},
$d = 2.5$\,nm, $a=0.44$\,nm, and $\theta_D=116$\,K, we obtain $\tau_{esc} \approx 120$\,ps.

\bibliography{mg_bibliography}

\begin{thebibliography}{102}
\bibitem [{EPH()}]{EPH_vs_eph}%
  \BibitemOpen
  \href@noop {} {}\bibinfo {note} {The dephasing time due to e-ph scattering is
  identical to the single-particle e-ph scattering time, $\tau_{e-ph}$.The e-ph
  energy relaxation time, measured by the AMAR method, is proportional to the
  $\tau_{e-ph}$ with a proportionality coefficient, which depends on the
  exponent $p$: it is about 0.6 for $p= 2.0$.}\BibitemShut {Stop}%
\end{thebibliography}

\end{document}